# Simulated spatial and temporal dependence of chromium concentration in pure Fe and Fe-14%Cr under high dpa ion irradiation

K. Vörtler[1], M. Mamivand[1], L. Barnard[1], I. Szlufarska[1], F. A. Garner[2], D. Morgan[1]

[1] Department of Materials Science and Engineering University of Wisconsin – Madison, WI 53706, USA
[2] Radiation Effects Consulting, Richland, WA 99354, USA
Corresponding Author: Dane Morgan, ddmorgan@wisc.edu, 608-265-5879

**Abstract**

In this work we develop an *ab initio* informed rate theory model to track the spatial and temporal evolution of implanted ions (Cr$^+$) in Fe and Fe-14%Cr during high dose irradiation. We focus on the influence of the specimen surface, the depth dependence of ion-induced damage, the damage rate, and the consequences of ion implantation, all of which influence the depth dependence of alloy composition evolving with continued irradiation. We investigate chemical segregation effects in the material by considering the diffusion of the irradiation-induced defects. Moreover, we explore how temperature, grain size, grain boundary sink strength, and defect production bias modify the resulting distribution of alloy composition. Our results show that the implanted ion profile can be quite different than the predicted SRIM implantation profile due to radiation enhanced transport and segregation.

## 1. Introduction

Charged particle irradiation has been used extensively for simulation of radiation damage induced in metals and alloys employed as structural components of nuclear reactors. These techniques have the advantage that they are cheaper, safer, and faster than high dose neutron irradiation experiments [1]. Typical charged particles used for simulation are high-energy electrons and protons, and metal or gas ions of various atomic weights. In the latter class, relatively light-weight ions such as He and neon are often used, and relatively heavy-weight ions such as Xe, Pt and Au have also been used. With the exception of He, however, these ions are not naturally produced in significant quantities during neutron irradiation. It is generally accepted that the best charged ions for simulation of neutron irradiation are those which are the major components of the alloy, primarily Fe, Ni and Cr for most nuclear steels, usually referred to as self-ions. Although Fe ions are the optimum choice for Fe-base alloys, Ni is frequently used and in some cases, Cr is used.

The injected ion induces a compositional change when it comes to rest, and while the compositional modification is often relatively small, the injected interstitial acting as a physical rather than chemical entity has been recognized as influencing the radiation damage, e.g., exerting a powerful suppressive effect on void nucleation and the post-transient swelling rate [2–6]. However, compositional effects of implanted ion can be important for high irradiation cases. For instance, in fusion and advanced fission reactors, such as sodium fast reactor, structural materials will experience extreme conditions of high irradiation doses, up to few hundred displacements per atom (dpa). In creating such a high damage dose with ion irradiation, the implanted ion compositional change can be important. For example, to reach 500 dpa in a region that avoids sampling the injected interstitial range, a peak dose of 900 dpa is needed for 1.8 MeV $Cr^+$ ions, this damage accumulation would add ~20% additional Cr just behind the peak damage depth (see Figure 1) [7]. While the injected Cr is expected to diffuse away from the peak somewhat during its introduction, the final expected injected Cr profile is not well established and it is known that much lesser changes in Cr content significantly influence void swelling of Fe and Fe-Cr alloys during neutron irradiation [8–10].

Additionally, a homogeneous Fe-Cr alloy is expected to redistribute its elemental components in response to various vacancy and interstitial mechanisms operating on gradients in dpa rate inherent in self-ion irradiation and on defect gradients near grain boundary and specimen surfaces [11]. Since the precipitation, swelling and other important properties of Fe-Cr alloys are sensitive to the Cr content, it is necessary to determine the spatial and temporal dependence of Cr along the ion range in order to fully understand the resultant properties, e.g., swelling vs. depth profiles.

In this study we have chosen to model irradiations using 1.8 MeV $Cr^+$ ions because a large number of experimental studies have been conducted at the Kharkov Institute of Physics and Technology in the Ukraine on a wide variety of austenitic, ferritic, ferritic-martensitic and oxide-dispersion-strengthened ferritic alloys, exploring doses of 50-600 dpa at a dpa rate of 1- 2 $x10^{-2}$ dpa/sec [7,12–15].

In the Fe-Cr system both Cr enrichment or depletion on grain boundaries has been observed for different alloys, temperatures and irradiation doses [16]. The underling segregation mechanisms governing these changes are not yet fully understood, although a competition of vacancy and interstitial transport mechanisms is very likely operating to produce these diverse results. Most previous modeling efforts in the Fe-Cr system were focused on radiation induced segregation (RIS) of Cr in Fe-Cr alloys involving segregation at grain boundaries experiencing proton or neutron irradiation [16–19] where there were no damage gradients or implanted Cr distributions.

There has been a significant body of experimental work on RIS in ion irradiated Fe-Cr based alloys (e.g., [7,12–15,18–21]) but it not clear from these works what happens to implanted ions in the entire damage region, particularly at very high doses where the implanted ions can significantly change composition. In general, it is not understood how the injected interstitials impact local alloy composition and RIS in the damage region. The only previous modeling work directly addressing this question of which we are aware is the recent study by Pechenkin *et al.* who studied the combined effect of damage profiles and implanted ion profiles in the Fe-Cr system using 1.8 MeV $Cr^+$ and 7 MeV $Ni^+$ ions [11]. Pechenkin *et al.* provided a valuable study of several complex commercial alloys and associated ion and RIS effects using similar models to those applied in this work. Here we extend such modeling to also address the influence of grain size, grain boundary (GB) type, varying temperature, and defect production bias. We also consider the effect of ballistic mixing in our modeling. In this work we use a somewhat different set of diffusivity parameters than Pechenkin *et al.* which have recently been developed through a combination of *ab initio* simulation and experimental validation [17,22]. However, as shown later the values used in our study and that of Pechenkin *et al.* appear to give similar results.

The rate theory model in this work was parameterized using first principles and was shown to reproduce well the experimentally observed RIS response at grain boundaries under different damage doses and temperatures [17]. As predicted by the first principle study of Choudhury et al. [22] Cr enrichment was the dominant RIS response. These findings correspond well with other Cr segregation in Fe-Cr alloys observations in literature [16,23]. The detail of RIS model and its experimentally validation has been discussed in Field et al. [17]. Cr enrichment is also consistent with crossover plot of Allen et al. [23] which maps out the boundary of Cr RIS sign (enrichment or depletion) based on alloy Cr contents and irradiation temperature.

The aim of the present paper is to provide insights into defect transport mechanisms and the resulting RIS during high dose iron irradiation. We construct a rate equation model considering damage and Cr implantation profiles of 1.8 MeV $Cr^+$ irradiating both Fe-14%Cr (Fe-X%Cr values are given in atomic percent, Fe-14at.%Cr corresponds to about Fe-13wt.%Cr) and pure Fe. We investigate grain sizes of 0.02-10μm and the effect of GB type by exploring different grain boundary sink strengths. We explore how specific mechanisms e.g. the effect of temperature, biased GB sinks and defect production bias, alter the alloy composition after irradiation.

## 2. Methods

The irradiation conditions we consider are 1.8 MeV $Cr^+$ irradiating pure Fe and Fe-14%Cr, with a peak damage of about 900 dpa, an irradiation time of 50,000 s (~14 hours), and a total Cr ion flux of $1.2 \times 10^{17}$ ions $m^{-2}$ $s^{-1}$. The total flux was derived by integrating the Cr fluence in Figure 1 and dividing it by the irradiation time. We assume a temperature of 450 °C unless otherwise noted. These irradiation conditions are similar to those encountered in the experiments conducted at the Kharkov Institute of Physics and Technology [7,12–15,21]. The damage and implanted Cr profiles were calculated using the Kinchin-Pease option in SRIM-2013 [24,25]. They are shown in Figure 1. The dose rate profile was obtained from the figure by dividing the damage profile by the irradiation time.

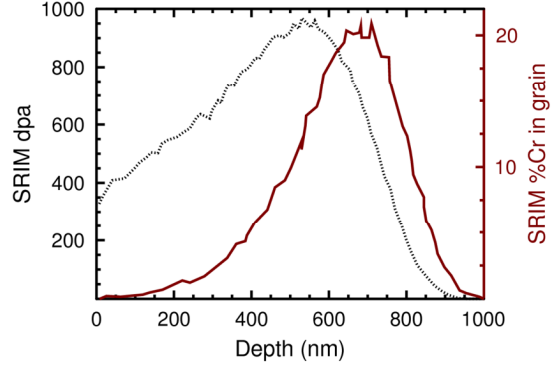

**Figure 1:** Damage and Cr implantation profiles as obtained by SRIM (Kinchin-Pease) **[24,25]**. The peak damage of 900 dpa is at about 500 nm depth, the peak of total implanted Cr (20 at.% Cr) at about 700nm depth.

The analysis done in this paper does not consider the formation of voids, self-interstitial loops, or the Cr rich $\alpha'$ phase, which is expected to start to be stable for concentrations of Cr near 8-12% at 450°C [27–29]. The formation of $\alpha'$ could significantly alter the quantitative compositions predicted here. While the formation of $\alpha'$ is likely under the Cr enriched concentrations predicted in this work it is unclear how the very high dose Cr irradiation might suppress or accelerate Cr precipitation. It was shown in overlapping cascades simulations without considering long time scale diffusion processes that cascades in Fe-15%Cr can form small Cr clusters [29]. However it is unclear if ion bombardment might also hinder bulk Cr-rich cluster formation due to ballistic remixing. Unfortunately, properly including the effects of void, loop, and $\alpha'$ formation and stability under high dose ion irradiation is beyond the scope of this work as it would involve a far more extensive modeling framework. Therefore, the reader should consider the Cr profiles in this work to be a qualitative guide to the trends and scales of effects that might be observed, rather than quantitative predictions.

A rate theory model of a two-component bcc alloy was employed [1,30]. The following coupled differential equations describing the site concentrations of Fe ($C_{Fe}$), Cr ($C_{Cr}$), vacancies ($C_v$), and self-interstitial atoms (SIAs) ($C_i$) depending on lattice plane were solved numerically:

$$\frac{\partial C_v}{\partial t} = -\nabla J_v + K_d(p)\varepsilon - RC_v C_i \qquad (1)$$

$$\frac{\partial C_i}{\partial t} = -\nabla J_i + K_d(p)\varepsilon - RC_vC_i + K_{Cr}(p) \tag{2}$$

$$\frac{\partial C_{Fe}}{\partial t} = -\nabla J_{Fe} + \frac{\partial C_{Fe}^{bal}}{\partial t} \tag{3}$$

$$\frac{\partial C_{Cr}}{\partial t} = -\nabla J_{Cr} + K_{Cr}(p) + \frac{\partial C_{Cr}^{bal}}{\partial t} \tag{4}$$

The equations (1) and (2) do not have the production bias and the annihilation terms on sinks which will be introduced later in equations (23) and (24). In above equations the $\nabla J_j$ describes the gradient of the flux of species $j$, $K_d(p)$ and $K_{Cr}(p)$ are the Frenkel pair and Cr implantation rates depending on lattice plane $p$ respectively, and $\varepsilon$ is the damage efficiency for heavy ions which is 30% approximately [31,32]. The Cr implantation rate was obtained from Figure 1 by dividing the final Cr percentage (depending on irradiation depth) by the irradiation time. $R$ is the recombination rate coefficient for vacancies and self-interstitials [30]:

$$R = 4\pi d_{rec} \frac{D_v + D_i}{V_{at}}, \tag{5}$$

where $V_{at}$ is the atomic volume ($V_{at} = a^3/2$ where $a$ is the lattice constant for bcc Fe), $D_v$ and $D_i$ are the diffusion coefficients for vacancies and SIAs, and $d_{rec}$ is the recombination distance, taken as $3.3a$ where $a$ is the lattice constant for Fe [33]. We use $a$ = 2.86 Å. As one can see in equations (1) and (2), we assume that the Cr bombardment produces the same number of SIAs and vacancies.

The terms $\frac{\partial C_{Cr}^{bal}}{\partial t}$ and $\frac{\partial C_{Fe}^{bal}}{\partial t}$ give the rate of change of concentration due to ballistic mixing for Cr and Fe, respectively. Following Martin [34], we can write the discrete version of the ballistic mixing terms as,

$$\frac{\partial c_i(p)}{\partial t} = \sum_{m=0}^{n} \gamma_i(m) b_i(m) [-2c_i(p) + c_i(p+m) + c_i(p-m)], \tag{6}$$

where $c_i(p)$ is the concentration of species $i$ at the plane $p$, $\gamma_i(m)$ is the ballistic jump frequency of the species $i$ at the plane $m$ (which is equal to dpa/s for that plane), $b_i(m)$ is the probability that a jump starting from planes $p \pm m$ would stop at plane $p$, and $n$ is the number of the planes in the ballistic mixing range. The ballistic mixing range is typically considered as a multiple of nearest neighbor distance in the crystal structure. Following Enrique and Bellon [35], we consider the ballistic mixing range to be $5a_{NN}$ (~12 Å) where $a_{NN}$ is the nearest neighbor distance in bcc Fe. $b_i(m)$ was specified based on normal distribution with a mean of zero and standard deviation equal to ballistic mixing range.

The diffusion coefficients of vacancies and SIAs are written as $D_v = d_v^{Fe}C_{Fe} + d_v^{Cr}C_{Cr}$ and $D_i = d_i^{Fe}C_{Fe} + d_i^{Cr}C_{Cr}$, respectively. Similarly, the diffusion coefficients for Fe and Cr are written $D_{Fe} = d_v^{Fe}C_v + d_i^{Fe}C_i$ and $D_{Cr} = d_v^{Cr}C_v + d_i^{Cr}C_i$, respectively, i.e. we assume that Fe and Cr migrate exclusively via vacancies and SIAs. The diffusivities $d_v^{Fe}$, $d_v^{Cr}$, $d_i^{Fe}$, and $d_i^{Cr}$ are taken from [22], and are based on *ab initio* calculations for dilute Fe-Cr. The diffusion coefficients and our general model approach have been used previously to model RIS in model Fe-9%Cr binary alloys and showed very good agreement with experiment [17]. The diffusivities are used in Arrhenius form as in [17] and prefactors and activation energy terms for the *d*-values are given in Table 1.

**Table 1:** Diffusivity pre-exponential factors and activation energies used in the rate theory model.

| Parameter | Notation | Value | Ref. |
|---|---|---|---|
| Pre-exponential factor for Fe SIA diffusivity | $d_{0,Fe}^{i}$ | 6.59 x 10$^{-7}$ m$^2$ s$^{-1}$ | [22] |
| Pre-exponential factor for Cr SIA diffusivity | $d_{0,Cr}^{i}$ | 6.85 x 10$^{-7}$ m$^2$ s$^{-1}$ | [22] |
| Pre-exponential factor for Fe vacancy diffusivity | $d_{0,Fe}^{v}$ | 5.92 x 10$^{-6}$ m$^2$ s$^{-1}$ | [22] |
| Pre-exponential factor for Cr vacancy diffusivity | $d_{0,Cr}^{v}$ | 5.46 x 10$^{-6}$ m$^2$ s$^{-1}$ | [22] |
| Activation energy for Fe SIA diffusivity | $E_a^{Fe,i}$ | 0.36 eV | [22] |
| Activation energy for Cr SIA diffusivity | $E_a^{Cr,i}$ | 0.27 eV | [22] |
| Activation energy for Fe vacancy diffusivity | $E_a^{Fe,v}$ | 0.77 eV | [22] |
| Activation energy for Cr vacancy diffusivity | $E_a^{Cr,v}$ | 0.68 eV | [22] |

The initial atomic Fe and Cr concentrations (given as site fraction) are set to 0.86 and 0.14, respectively, for calculations in Fe-14%Cr. We assume that the initial (impurity) atomic Cr concentration in the pure Fe sample is zero, which we take as 10$^{-20}$ (Fe impurity site fraction) for numerical purposes. The initial vacancy and SIA concentrations are defined as

$$C_k = e^{-\frac{E_k^f}{k_B T}}, \quad (7)$$

where $C_k$ is the defect concentration, $E_k^f$ the formation energy of defect $k$ at temperature $T$, and $k_B$ the Boltzmann constant. In this work we use the formation energy parameters as in Ref. [17].

The equations are discretized and solved numerically as in [17,36]. The gradient of the fluxes for species $j$ in the rate equations become finite difference equations

$$\frac{dJ_j^n}{dx} = -\left[\frac{J_j^{n+1} - J_j^n}{\Delta x}\right], \quad (8)$$

where $n$ is the plane number and $\Delta x$ the distance between lattice planes. According to [1] the fluxes are calculated as

$$J_{Fe}^n = -D_{Fe}\frac{C_{Fe}^n - C_{Fe}^{n-1}}{\Delta x} - C_{Fe}^n \frac{d_i^{Fe}(C_i^n - C_i^{n-1}) - d_v^{Fe}(C_v^n - C_v^{n-1})}{\Delta x} \quad (9)$$

$$J_{Cr}^n = -D_{Cr}\frac{C_{Cr}^n - C_{Cr}^{n-1}}{\Delta x} - C_{Cr}^n \frac{d_i^{Cr}(C_i^n - C_i^{n-1}) - d_v^{Cr}(C_v^n - C_v^{n-1})}{\Delta x} \quad (10)$$

$$J_v^n = -D_v\frac{C_v^n - C_v^{n-1}}{\Delta x} + (d_v^{Fe} - d_v^{Cr})C_v^n \frac{(C_{Fe}^n - C_{Fe}^{n-1})}{\Delta x} \quad (11)$$

$$J_i^n = -D_i\frac{C_i^n - C_i^{n-1}}{\Delta x} - (d_i^{Fe} - d_i^{Cr})C_i^n \frac{(C_{Fe}^n - C_{Fe}^{n-1})}{\Delta x} \quad (12)$$

In the case of equal numbers of vacancy and interstitial defect fluxes, the fluxes fulfill in each plane

$$J_{Fe}^n + J_{Cr}^n = -J_v^n + J_i^n, \quad (13)$$

meaning that they are balanced.

However, in some model cases, the fluxes of interstitials and vacancies will not be balanced, and the condition in equation (13) will not hold. Due to these unequal fluxes of vacancies and SIAs to the boundary plane mass might not be conserved on the surface. Numerically this mass loss is compensated by the parameter $K_{loss}$ (see discussion in [36]). Thus, the boundary conditions of the rate equations are as in [36] and the parameter $K_{loss} = \frac{dC_{Fe}^1}{dt} + \frac{dC_{Cr}^1}{dt}$ is defined. At the boundary plane 1 mass balance is restored by subtracting a negative term $K_{loss}$. The Fe and Cr concentrations in the first plane are redefined as:

$$\frac{dC_{Fe}^1}{dt} \rightarrow \frac{dC_{Fe}^1}{dt} - C_{Fe}^1 K_{loss} \qquad (14)$$

$$\frac{dC_{Cr}^1}{dt} \rightarrow \frac{dC_{Cr}^1}{dt} - C_{Cr}^1 K_{loss} \qquad (15)$$

$$\frac{dC_v^1}{dt} = \frac{dC_i^1}{dt} = 0 \qquad (16)$$

Moreover, at the boundary planes 1 and $n_{max}$ the fluxes of all species *j*: vacancy, SIA, Cr, and Fe concentrations are kept constant (we assume the system is symmetric about the grain center, so the flux is zero at both those planes):

$$\frac{dJ_j^1}{dt} = 0; \frac{dJ_j^{n_{max}}}{dt} = 0. \qquad (17)$$

The size of the simulation cell is 500,000 lattice planes (101.1µm). For the calculation with multiple GBs a loss of vacancies and SIAs at the grain boundaries is assumed. The GB is assumed to be of low angle (consistent with the MA957 alloy used in the experiments of Ref. [7,37]) that can be described as an array of dislocations. Therefore, we use the definition of dislocation sinks following Ref. [36], which allows us to treat GBs as a locally high dislocation density. We therefore introduce dislocation density into the model as $\rho_{GB}(p)$, which is not constant but depends on the lattice plane *p*. The rate equations for vacancies and SIAs (equations (1) and (2)) then become

$$\frac{\partial C_v}{\partial t} = -\nabla J_v + K_d(p)\varepsilon - RC_vC_i - GB_v(p) - DL_v \qquad (18)$$

$$\frac{\partial C_i}{\partial t} = -\nabla J_i + K_d(p)\varepsilon - RC_vC_i + K_{Cr}(p) - GB_i(p) - DL_i, \qquad (19)$$

where the grain boundary sinks for species *k* (vacancies and SIAs) are defined as

$$GB_k(p) = Z_k^{GB} 4\pi r_k D_k \frac{\rho_{GB}(p)}{\Delta x} C_k, \qquad (20)$$

where $Z_k^{GB}$ is the GB bias factor for vacancies or SIAs that is taken as 1.0 unless otherwise stated. The radius of capture for defect species k is $r_k$, which is approximately few times of lattice parameter and we consider it to be $1\ nm$. The concentration and diffusion coefficient for defect $k$ are $C_k$ and $D_k$, respectively. $\rho_{GB}(p)$ is the density of dislocations associated with the GB and is given by

$$\rho_{GB}(p) = \begin{cases} \rho_{gb}; & for\ [l_p^m \ldots l_q^m] \\ 0; & else \end{cases}, \qquad (21)$$

where $\rho_{gb}$ is set to $2.8 \times 10^{17}\ m^{-2}$ corresponding to an approximately 7.5 degree low angle GB unless otherwise specified. The relationship between GB dislocation density and grain boundary angle was determined using $d = b/\theta$ [38], where $d$ is the dislocation spacing, $b$ the Burger's vector and $\theta$ the GB angle in rad. We assume a <110>{111}-dislocation slip system. $l_c^m$ represents the GB of number $m$ and in planes $p$ to $q$. A GB sink width of q-p=3 lattice planes is used. Grains boundaries are assumed to be spaced uniformly along the direction normal to the surface. This geometry is idealized but provides a qualitative representation of how different grain boundary concentrations and strengths couple to the radiation induced defect concentrations and associated chemical changes. We assume the effect of a constant "background" dislocation density (the loss of vacancies and SIAs to dislocations) can be written for a defect $k$ as

$$DL_k = 4\pi r_k D_k \frac{\rho_{DL}}{\Delta x} C_k, \qquad (22)$$

where $\rho_{DL}$ is the dislocation density which is set to $10^{14}\ m^{-2}$ (based on values typically seen in high-dose irradiations of Fe-Cr steels, e.g., see Ref. [39]) unless otherwise specified. $D_k$, $r_k$, and $C_k$ are the diffusion coefficient, radius of capture, and concentration of defect k, respectively.

To simulate a defect production bias, mimicking the effect of immobile clusters that result in unequal numbers of vacancies and SIAs at high ion energies, a bias factor $Z_k^{FP}$ is added to the source term in the rate equations for vacancies and SIAs. Thus equations (17) and (18) become

$$\frac{\partial C_v}{\partial t} = -\nabla J_v + Z_v^{FP} K_d(p)\varepsilon - RC_v C_i - GB_v(p) - DL_v \qquad (23)$$

$$\frac{\partial C_i}{\partial t} = -\nabla J_i + Z_i^{FP} K_d(p)\varepsilon - RC_v C_i + K_{Cr}(p) - GB_i(p) - DL_i. \qquad (24)$$

These bias terms are both set to one (no bias) unless mentioned explicitly.

The Cr and Fe concentrations have the units of atom fraction per initial lattice site. Due to the source term of Cr from the implanted ions in equation (2) and due to the mass loss by unequal fluxes of vacancies and SIAs to GB sinks, after time evolving the system the sum of Cr and Fe concentrations in a lattice site may not be 1.0. Therefore, after the simulation has run the calculated $C_{Cr}$ is renormalized to give the Cr fraction in each plane to be the final Cr atomic fraction as follows:

$$Cr\ fraction = \frac{N_{Cr}}{N_{Cr} + N_{Fe}} = \frac{C_{Cr}}{C_{Cr} + C_{Fe}} \tag{25}$$

where all values are in a given plane and $N_X$ is the total number of atoms of species $X$ in a plane.

## 3. Results

### 3.1. Local alloy composition during and after irradiation

Figure 2 shows the Cr content depending on depth for Fe-14%Cr and pure Fe after 1.8 MeV Cr$^+$ irradiation to a 900 dpa peak dose. One can see that both the large grain ("large grain" refers to a grain that is enough larger than the damage zone that the grain boundary has no impact on the results) and a 0.3 µm grain alloys have Cr enrichment at the ion-incident surface. We also compare our results to those of Pechenkin *et al.* [11], and find only a small difference in the Cr profile. This result is consistent with the fact that although Pechenkin *et al.* used somewhat different defect diffusivities than this work, the overall values are similar and the ratio difference which drives RIS, $d_{Cr}^i/d_{Fe}^i - d_{Cr}^v/d_{Fe}^v$, is quite close.

Differences to the large grain profile can be seen comparing it to the small grain simulation: following the implantation profile there is a spread in the Cr composition with RIS at the GB. The effect of the implanted Cr is shown in the large grain "damage only" profile, which was obtained by removing the Cr ion implantation profile from the rate equations. This "damage only" profile removes the implanted ions and is therefore more consistent with what one might get from protons rather than Cr irradiation.

The time dependence (i.e. dose dependence using a constant dose rate) of the Cr profiles of 0.3 µm grains for both Fe-14%Cr alloy and Fe are shown in Figure 3. No saturation of the Cr fraction depending on implantation time can be seen, i.e. no steady state of Cr segregation via defect diffusion and Cr implantation is reached before the peak dose of 900 dpa is reached after 50,000s. This result is due to the constant addition of implanted Cr.

Moreover, one can see that the evolving local Cr compositions in the alloy are quite different than predicted by the SRIM profiles. The effects shown in Figure 2 are explored in the following sections.

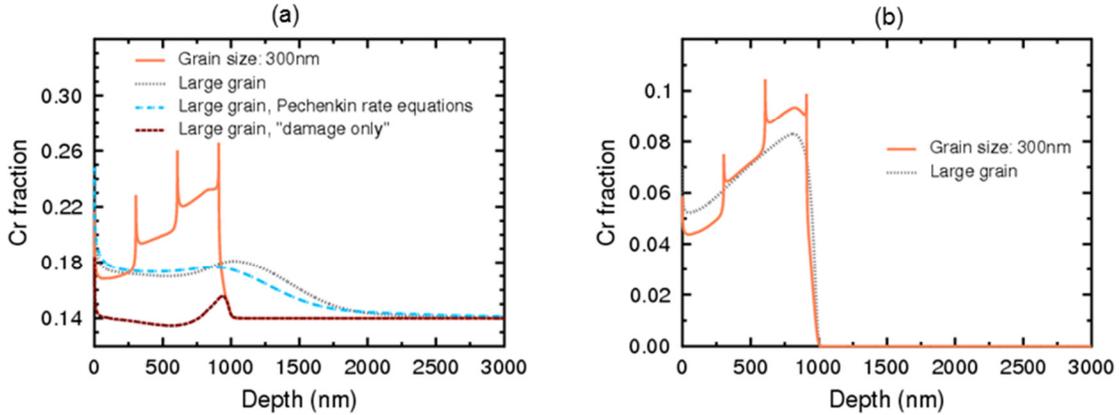

**Figure 2:** Cr fraction vs. depth (in nm) for Fe-14%Cr (a) and pure Fe (b). Left figure also shows Pechenkin equations **[11]** with $k^2 = 10^{12}$ m$^{-2}$ sink strength as defined in **[11]** which corresponds to $\rho_{DL} = 6\times10^{10}$ m$^{-2}$. The "large grain" refers to one that is enough larger than the damage zone that the grain boundary has no impact on the results (here the large grain diameter is set to 3,000nm).

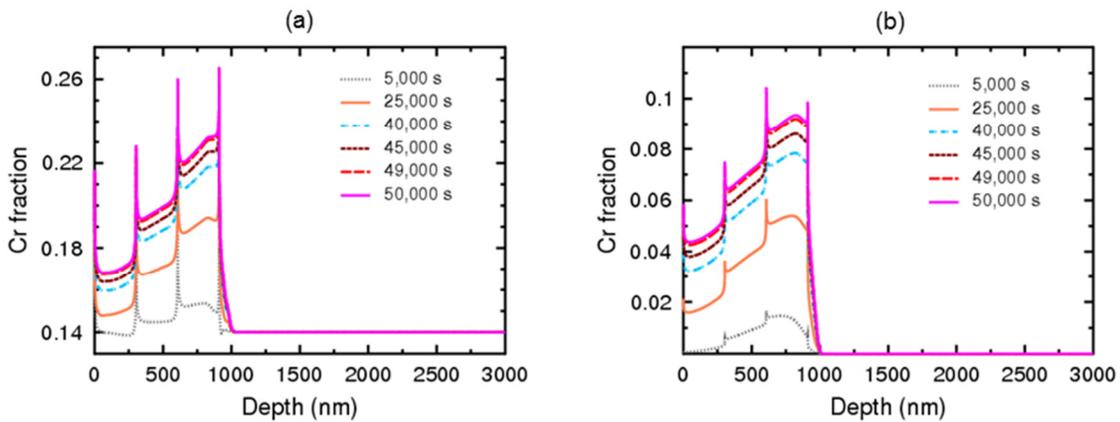

**Figure 3:** Time dependence of Cr fraction vs. depth (in nm) from 5,000 s to 50,000 s for Fe-14%Cr (a) and pure Fe (b) with 300 nm grain size. The peak dose at 50,000 s is about 900 dpa.

*3.2. Effect of grain size*

Figure 4 shows the effect of grain size on Cr concentration for Fe-14%Cr (Figure 4(a)), pure Fe (Figure 4(b)) under $Cr^+$ ion irradiation and for Fe-14%Cr (Figure 4(c)) under "damage only" irradiation with a peak dose of 900 dpa. As one can see in the figure, the local alloy composition as a function of irradiation depth changes depending on the grain size. There is some RIS, showing Cr enrichment, near all the grain boundaries. In the ion irradiated Fe and Fe-14%Cr the Cr profiles have a shape with some similarity to the Cr SRIM implantation profile, in that they peak around 700nm and end near 1000nm. However, transport to the surface is clearly moving significant Cr to the surface, and the Cr content near the surface is generally higher than in the original alloy or that expected from just the SRIM profile which the changes in Cr with depth are smaller than the SRIM profile. The similarity between the final Cr concentration and the SRIM profile is due to the high dislocation and GB concentration that reduces the defect concentrations and therefore reduces the mechanism of Cr redistribution. The implanted Cr is therefore somewhat stable over the time of the experiments. The "damage only" profile in Fe-14%Cr shows that it follows the broad "damage only" profile in Figure 2 however, there is Cr enhancement due to RIS on the GBs. In both Cr irradiated Fe and Fe-14%Cr the larger sink strength of the small grain compared to large grain cases causes the Cr profiles follow more closely the SRIM Cr implantation profile in the small grain compared to large grain cases. The results show that for all grain boundary sizes, no Cr segregation survives after 1000 nm, which is the end of damage and Cr implantation depth.

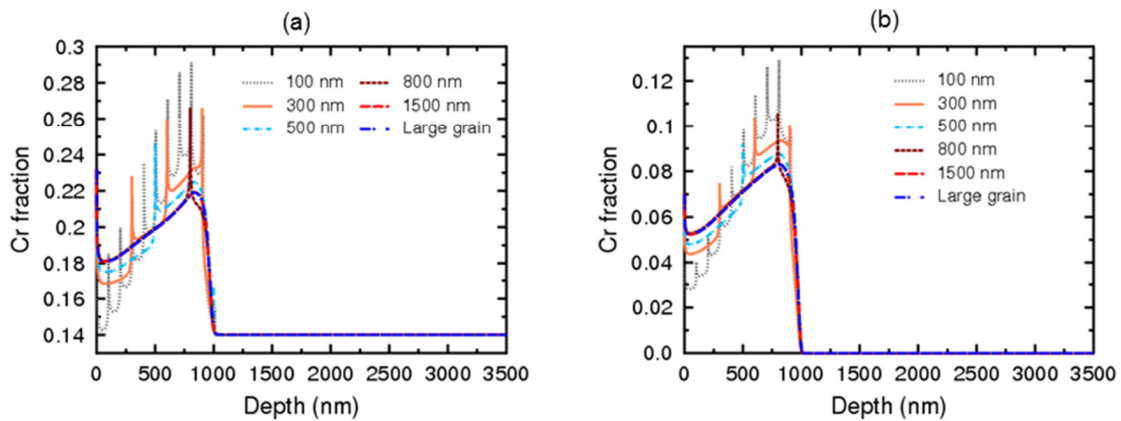

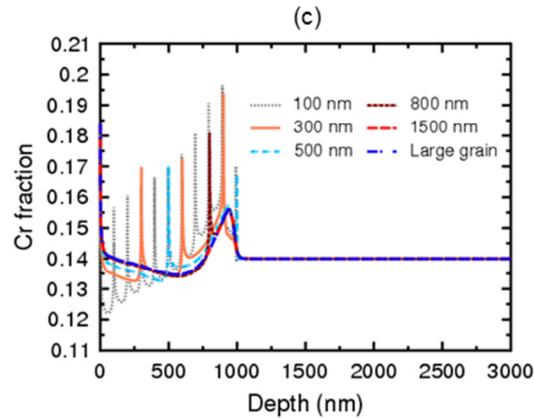

**Figure 4:** Cr fraction depending on depth in nm: (a) Fe-14%Cr Cr-ion irradiation, (b) pure Fe Cr-ion irradiation, and (c) Fe-14%Cr "damage only".

In Figure 5(b) the change in Cr GB RIS is shown for 20 nm, 50 nm, 100 nm, 200 nm, and 400 nm grain in a "damage only" irradiation condition with a peak dose of 900 dpa. The GB segregation is compared with both Cr bulk concentration (14%) and Cr local concentration at the valley of Cr concentration profile. Measurements are taken at 400 nm depth. One can see the smallest change in RIS is for the 20 nm case and that the change in RIS is approximately linear, depending on grain size up to 100 nm grain size. For grain sizes larger than 100 nm the Cr GB RIS is almost constant. The corresponding Cr profile is given in Figure 5(a). This result shows that the RIS decreases significantly with decreasing grain size. This effect may be due both to interactions between the grains associated with overlapping or nearly overlapping Cr enrichment profiles and the generally increased sink density that occurs with smaller grains.

The RIS model in this work has already been benchmarked against experimental results [17]. In addition, the predicted amount of Cr segregation is in agreement with other experimental data. For instance, Allen et al. [23] reported 5 at.% enrichment for Cr in 14YWT ODS alloy which was irradiated to 100 dpa with 5.0 MeV $Ni^{++}$ ion and Marquis et al. [40] reported approximately 5 at.% Cr segregation for Fe-12Cr irradiated with 0.5-2.0 Mev Fe ion at 500 °C up to $1\text{-}13 \times 10^{15}$ /$cm^2$. All these experimental results are in good agreement with model prediction of 4 at.% Cr segregation for large grains.

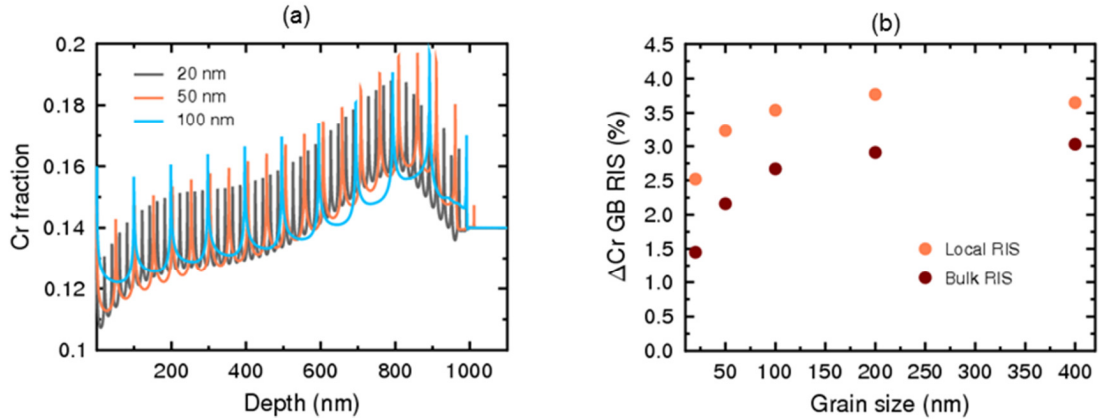

**Figure 5:** (a) Cr fraction vs. depth for 20 nm, 50 nm and 100 nm grain size for "damage only" irradiation conditions in Fe-14%Cr. (b) shows the corresponding **ΔCr GB RIS** in comparison with bulk concentration and local concentration (the approximate change in Cr concentration between the GB peak and valley). The measurements are taken at 400 nm depth.

*3.3. Effect of sink strength on grain boundaries and dislocations*

Changing the parameter $\rho_{gb}$ in equation (21) can be used to explore the effect of different GB types because the GB structures associated with high angle GBs or low angle GBs are different and this alters their ability to capture vacancies and SIAs. The effect of parameter $\rho_{gb}$ from equation (21), the GB sink strength, is shown in Figure 6. In the figure is shown a Cr profile of a 300 nm grain using different sink strengths, and it can be seen that the profile flattens out as the sink becomes weaker . For GB sink strength lower than $\rho_{gb} = 10^{15} \ m^{-2}$, a GB sink has almost no effect compared to using no GB sink or simulating a very large grain. For weak GBs the Cr concentration is ascending between GBs intervals while for strong GBs the Cr concentration remains almost constant.

We note that in this model the GBs are assumed to be parallel to the sample surface. In real materials there is possibility that grains be at any angle to the surface. However, modeling GBs at angles that are not parallel to the surface requires expanding the model to at least two-dimensions, which is beyond the scope of this work. However, we do not believe that the introduction of other GB orientations will make a significant impact on the results. In the current model the grain boundaries and surfaces act as sinks which are connected by effectively infinitely fast diffusion paths. This assumption is common in RIS models and is justified by the fact that the GB diffusion is significantly faster than the bulk volume diffusion. Therefore, all the GBs effectively communicate with each other and the surface. Adding differently oriented GBs will alter the average sink density and the locations of the RIS, but should have minimal impact on the qualitative shape of the Cr profile.

We also note that within our model we treat the surface as in equilibrium with the GBs and assume that both surface and GBs have the same energetics. However, if Cr is significantly more stable on the surface than in GBs Cr would move from the GBs to the surface, which could effectively deplete Cr from bulk of the material. It is beyond the scope of the present paper to explore this effect but we can estimate the range of grain sizes over which it might play a role by considering the diffuse length of Cr in a GB under the experimental conditions.

From Radis et al. [41] we know the ratio of iron self-diffusion in grain boundary to bulk at 450 °C is about $10^7$. It is also a well-accepted assumption that the ratio $D_{GB}/D_{bulk}$ in Fe is similar between impurities and self-diffusion [41]. Therefore, at 450 °C we estimate $D_{Cr}^{GB} \approx 10^7 D_{Cr}^{Bulk} = 2.26 \times 10^{-18} m^2/s$ and the diffusion length of Cr in a GB during the experiment time (14hr) is $L = \sqrt{D \times t} = 330 nm$. Therefore, we can conclude that grain boundary and surface intersection might change the Cr profiles for materials with grain size smaller than about 330 nm.

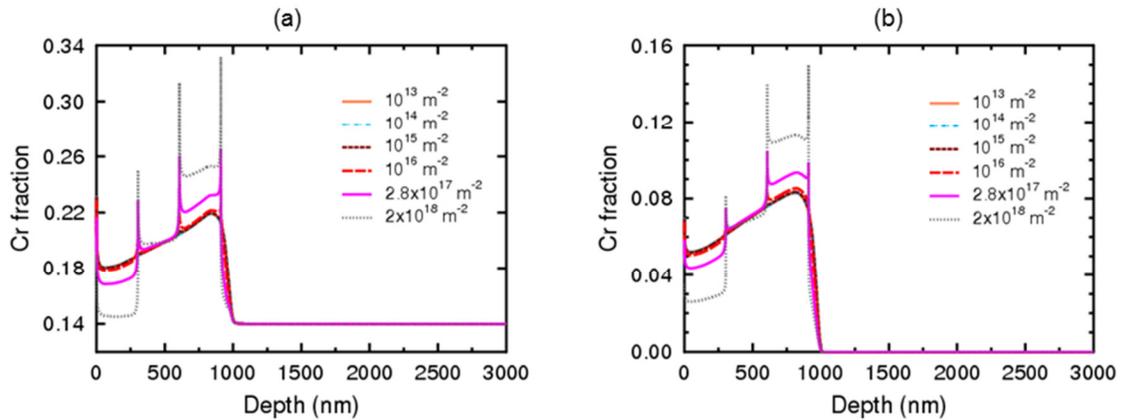

**Figure 6:** Cr fraction vs. depth in nm in Fe-14%Cr (a) and pure Fe (b) for different GB sink strength parameters $\rho_{gb}$. The peak dose is 900 dpa.

In Figure 7 we show the effect of the background dislocation sink strength $\rho_{DL}$ (as in equation (22)). One can see in the figure that at sink strengths smaller than $\rho_{DL} \sim 10^{10} \ m^{-2}$ there is little effect on the Cr profile compared to case without dislocations. For very high dislocation densities, the implanted Cr profile begins to look similar to the nominal SRIM profile, as the high density dislocations removes most point defects before they can enable Cr diffusion away from the peak region. Figure 7 also shows that the Cr profile remains almost unchanged for background dislocation densities higher than $10^{16} \ m^{-2}$ and that the profile is within about 3% (Fe) – 6% (Fe-14Cr) of that predicted by the SRIM simulation. In general, from Figure 4, Figure 6, and Figure 7 we see that increasing dislocation sink density, either through grain boundary number, grain boundary strength, or background dislocation density, respectively, leads to a similar trend of reducing defects and retaining more of the initial SRIM profile.

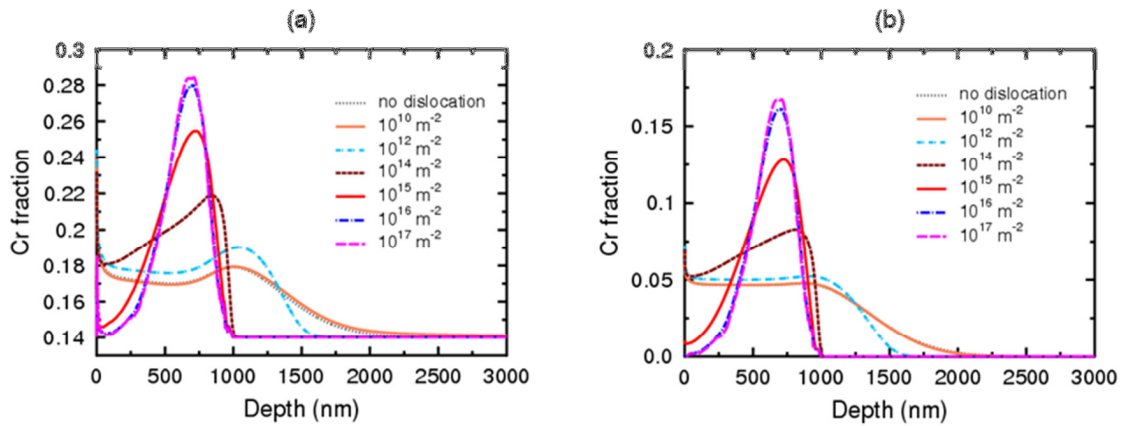

**Figure 7:** Cr fraction vs. depth in Fe-14%Cr (a) and pure Fe (b) for different background dislocation sink strength parameters $\boldsymbol{\rho_{DL}}$.

*3.4. Effect of temperature on Cr concentration profiles*

When the temperature in the Fe-Cr alloys increases, there are both increases and changes in the ratios of the Cr and Fe SIA and vacancy diffusivities. Moreover, initially more vacancies are present in a bulk material for higher temperatures, although the faster defect diffusion at the higher temperature is the dominant effect. Figure 8 shows the effect of the temperature for a large grain simulation. For both Fe-14%Cr and pure Fe the temperature does not have a noticeable effect on Cr concentration profile. The Cr profile is similar to that seen in the predicted implantation profile in the SRIM simulation (Figure 1). The temperature effect on Cr profile is minor because of high background dislocation density ($10^{14}\ m^{-2}$), which captures the majority of defects and minimizes the effect of temperature on changing the SIA and vacancies diffusivity.

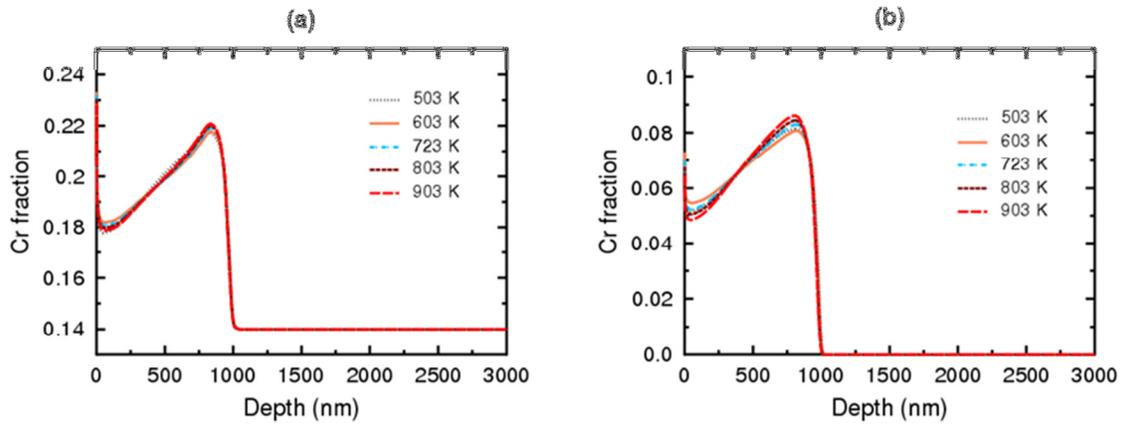

**Figure 8:** Cr fraction vs. depth in nm depending on temperature: (a) Fe-14%Cr and (b) pure Fe.

*3.5. Effect of defect production bias*

High energy ion irradiation of Fe-Cr alloys produces not only single vacancies and SIAs, but also large immobile defect clusters [42]. Consequently, the number of mobile SIAs and vacancies can be different. The effect of a defect production bias is shown in Figure 9 for one large grain. In all cases a peak dose of 900 dpa was used. In the figure a vacancy bias of 0.9 means that 10% of the produced vacancies are immobile, i.e. they are trapped in voids and a SIA bias means a corresponding trapping of interstitials.

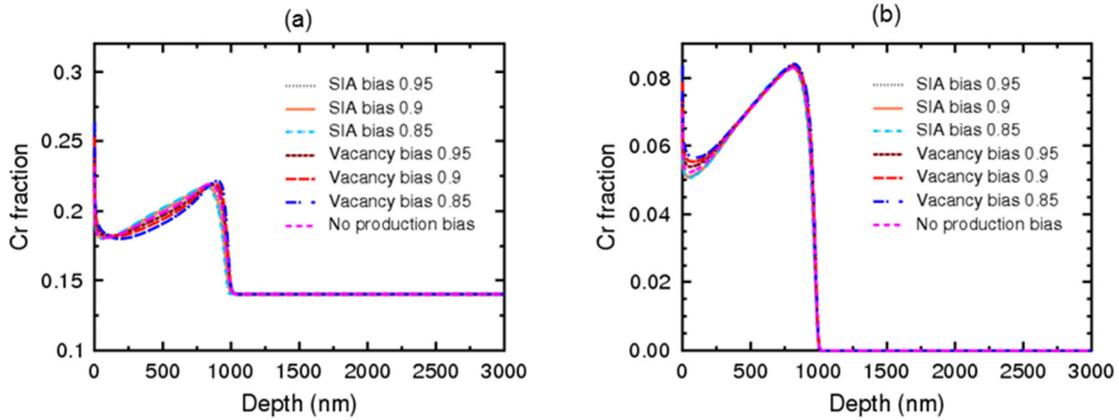

**Figure 9:** Cr fraction depending on depth in nm in Fe-14%Cr (a) and pure Fe (b) for different defect production bias factors.

The defect bias values seen in Figure 9 are intended to show the general effect on the Cr profiles, they are not based on specific defect biases obtained by experiments. From overlapping cascade simulations the typical SIA clustered fraction is ~80% and vacancy clustered fraction is~50% in Fe-Cr alloys [29]. We assume that after diffusion of the smaller defect clusters produced by cascades and corresponding recombination 5-15% of the produced vacancies/SIAs are immobile. The results in Figure 9 demonstrate that the ratio of mobile defects (vacancies and SIAs) does not have a significant effect on the Cr segregation.

*3.6. Effect of biased grain boundary sinks*

The bias factor describes the preferential diffusion of SIAs (vacancies) relative to the diffusion of vacancies (SIAs) to the GB. Such a bias can alter the ratio of mobile vacancies and SIAs as more SIAs or vacancies get trapped at the GBs. In our modeling approach the GB sink bias is due to a dislocation bias [43] since the (low angle) GB is defined as an array of dislocations. The phenomenon is explored in Figure 10, which shows a 300 nm grain with a GB sink strength of $\rho_{gb} = 2.8 \times 10^{17}\ m^{-2}$ employing different vacancy and SIA GB biases. In the figure a vacancy (SIA) sink bias of 0.6 means that 60% of vacancies (SIAs) and 100% of SIAs (vacancies) are lost at GBs, and therefore 40% more SIAs (vacancies) than vacancies (SIAs) are captured at the GBs. Other GB sink bias values have a similar meaning. The results show Cr enriching in all grain boundaries within the damage and Cr implantation range. No enrichment is seen beyond 1000 nm, as found in the previous cases. Overall we find that GB sink bias does not have a significant effect in Cr segregation profile.

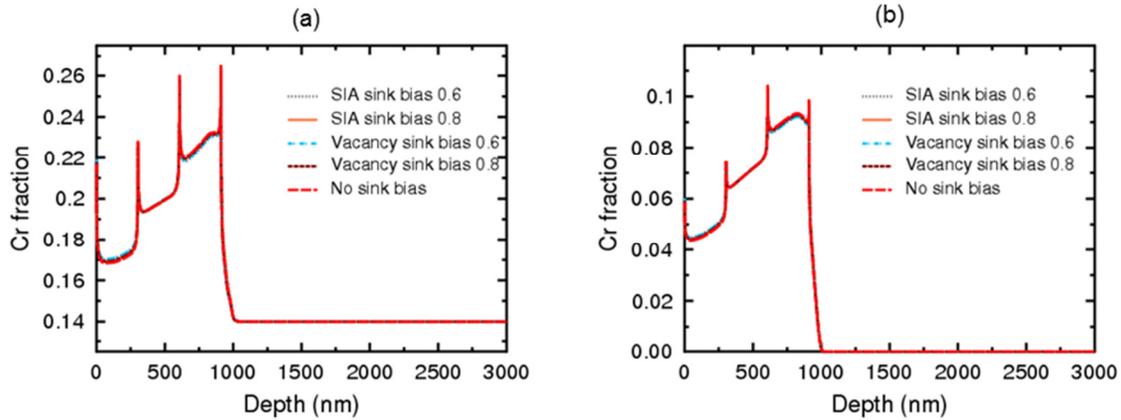

**Figure 10:** Cr fraction depending on depth in nm in Fe-14%Cr (a) and pure Fe (b): Vacancy sink bias 0.6 means that 60% of vacancies and 100% of SIAs (efficiency of sink) are lost at the GB: 40% more SIAs than vacancies are captured at the GB.

*3.7. Effect of ballistic mixing*

As an ion penetrates in a solid, it slows down and transfers its energy to the atoms and electrons of the solid. Ion collision with target atoms can displace them permanently from their lattice sites and relocate them several lattice sites away, causing a so-called "ballistic" mixing which is distinct from the mixing driven on longer time scales by the point defects created during the ion collision. Experimentally it has been observed that ballistic mixing can lead to interface mixing at the boundary of two different materials [44,45]. We added ballistic mixing to the model to capture its possible effects on widening sharp Cr segregation profile at grain boundaries. The details of the ballistic mixing model are described in section 2. Figure 11 shows the Cr profile for two cases: 1) with ballistic mixing effects in the model and 2) without ballistic mixing effects in the model. The results reveal that the ballistic mixing does not have noticeable effect on Cr deposition profile and the RIS yields quite similar profiles with and without ballistic mixing,

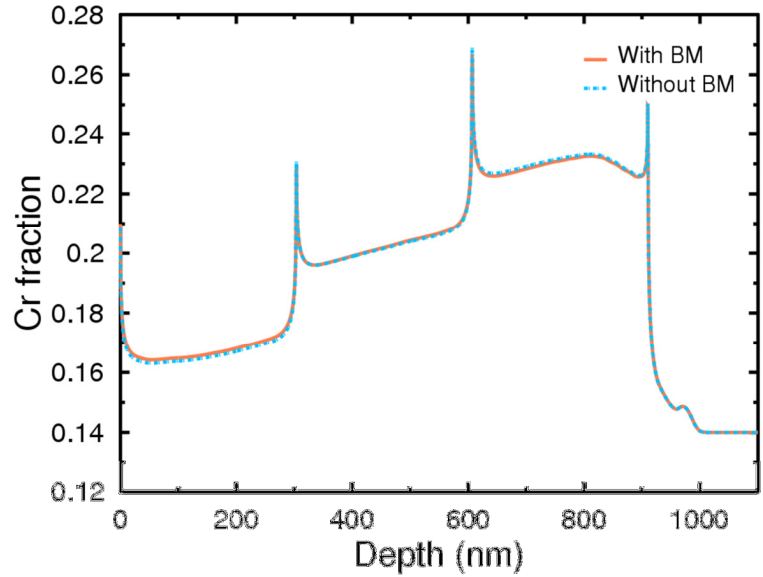

**Figure 11:** The effects of ballistic mixing in spreading the Cr profile are minor and the RIS is the dominant (Fe-14%Cr).

*3.8. Uniform irradiation*

We also use a uniform irradiation with no ion implantation to simulate the neutron irradiation condition for comparison to the ion irradiation conditions included in the rest of the paper. We consider dose rate of $2 \times 10^{-3}$ dpa/s and the same irradiation time as above (50,000s), yielding a total of 100 dpa uniform damage in the material. Figure 12 shows a flat profile (because of no ion implantation) with some segregated regions at the grain boundaries.

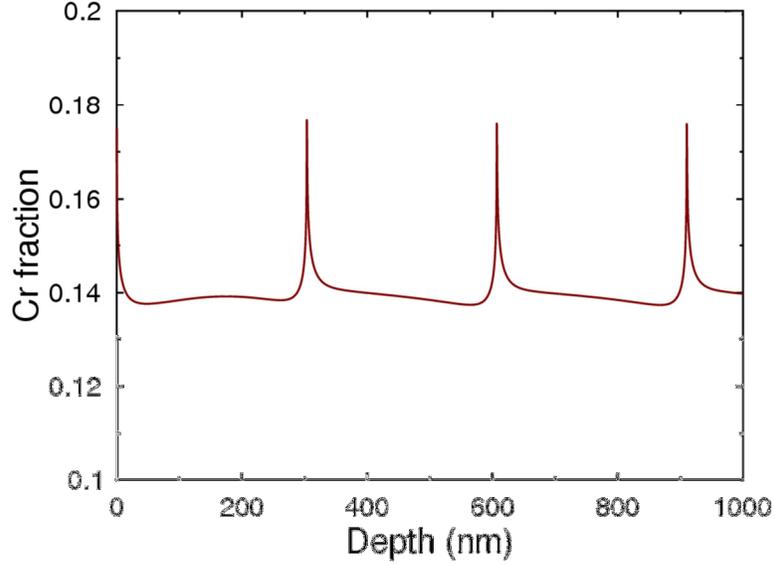

**Figure 12:** Cr distribution in Fe-14%Cr under uniform irradiation. Total deposited damage is 100 dpa.

## 4. Discussion

To better understand the origin of the trends shown in the Results section the vacancy and SIA concentration profiles at the end of the implantation for pure Fe are given in Figure 13 and Figure 14. Figure 13 shows the defect concentration profiles for different defect production bias factors (a),(b), temperatures (c),(d), and GB sink bias factors (e),(f), and Figure 14 illustrates defect concentration profiles for different GB sink strength parameters $\rho_{gb}$ (a), (b), grain sizes (c),(d), and background dislocation sink strength parameters $\rho_{DL}$ (e),(f). In the figures, one can clearly see the GB sinks for vacancies and SIAs, characterized by the kinks in the defect concentration profiles (e) and (f).

The vacancy and SIA concentration profiles for Fe-14%Cr have a similar shape as shown for pure Fe in Figure 13(a) to (f), however, the concentrations are smaller (up to about half as large as for pure Fe). Thus, in pure Fe the defect concentration is larger than in Fe-14%Cr.

As shown in Figure 13(c) and (d) the defect concentrations depend on temperature. Applying to equation (7) the vacancy formation energy in Fe $E_v^{fe}$ =2.1 eV at the experimental temperature 723 K, one gets the initial vacancy fraction per lattice site $C_{0v} = 2.3 \times 10^{-15}$. The mean diffusion distance for vacancies can be estimated as $r_{m_v} \approx 2\sqrt{D_v t_m}$. One can solve for the mean diffusion time

$$t_m \approx \frac{r_{m_v}^2}{4D_v}. \tag{26}$$

For vacancy mediated Cr diffusion the diffusion coefficient can be written as $D_v^{Cr} = C_v d_v^{Cr}$, where $C_v$ is the vacancy fraction per lattice site and $d_v^{Cr}$ the vacancy diffusivity for Cr in dilute FeCr ($d_v^{Cr} = 9.9 \times 10^{-11}\ m^2 s^{-1}$ at 723 K). Using equation (26) and assuming Cr can diffuse only by the initial vacancies in the material and $r_{m\_v} = 700\ nm$ (peak of Cr implantation profile in Figure 1) then $t_m$ is about $10^{11}$ seconds at the implantation temperature (much longer than the implantation time). We also note that during the experimental time scale of 14h the Cr is expected to diffuse under 10nm under thermal equilibrium conditions. This means that Cr diffuses almost entirely by radiation enhanced diffusion (mediated by the SIA and vacancies as a result of the heavy ion irradiation).

To assess if the Cr diffusion to the boundaries is mediated primarily by SIA or vacancies, one can determine the Cr to Fe diffusivity ratios. At 723 K the diffusivity ratio for vacancies is $\frac{d_v^{Cr}}{d_v^{Fe}} = 3.9$ and for SIAs is $\frac{d_i^{Cr}}{d_i^{Fe}} = 4.4$. Consequently, there is a slight enrichment of Cr at sinks due to the preferential Cr diffusion via SIAs.

We want to add that the diffusivities from Ref. [22] are only calculated for a dilute case, i.e. when the Cr concentration is very low. Recent studies on Ni-Cr alloys have shown that the composition dependence of interstitial diffusion can be quite significant and dramatically alter RIS predictions [46]. Similarly, work by Wharry et al. [16] has suggested a compositional dependence in diffusion coefficients in Fe-Cr. There is therefore a significant uncertainty in the diffusion parameters, especially for concentrated Fe-14%Cr. More accurate models can likely be formulated with properly concentration-dependent diffusion coefficients, but further work in this area is beyond the scope of this study.

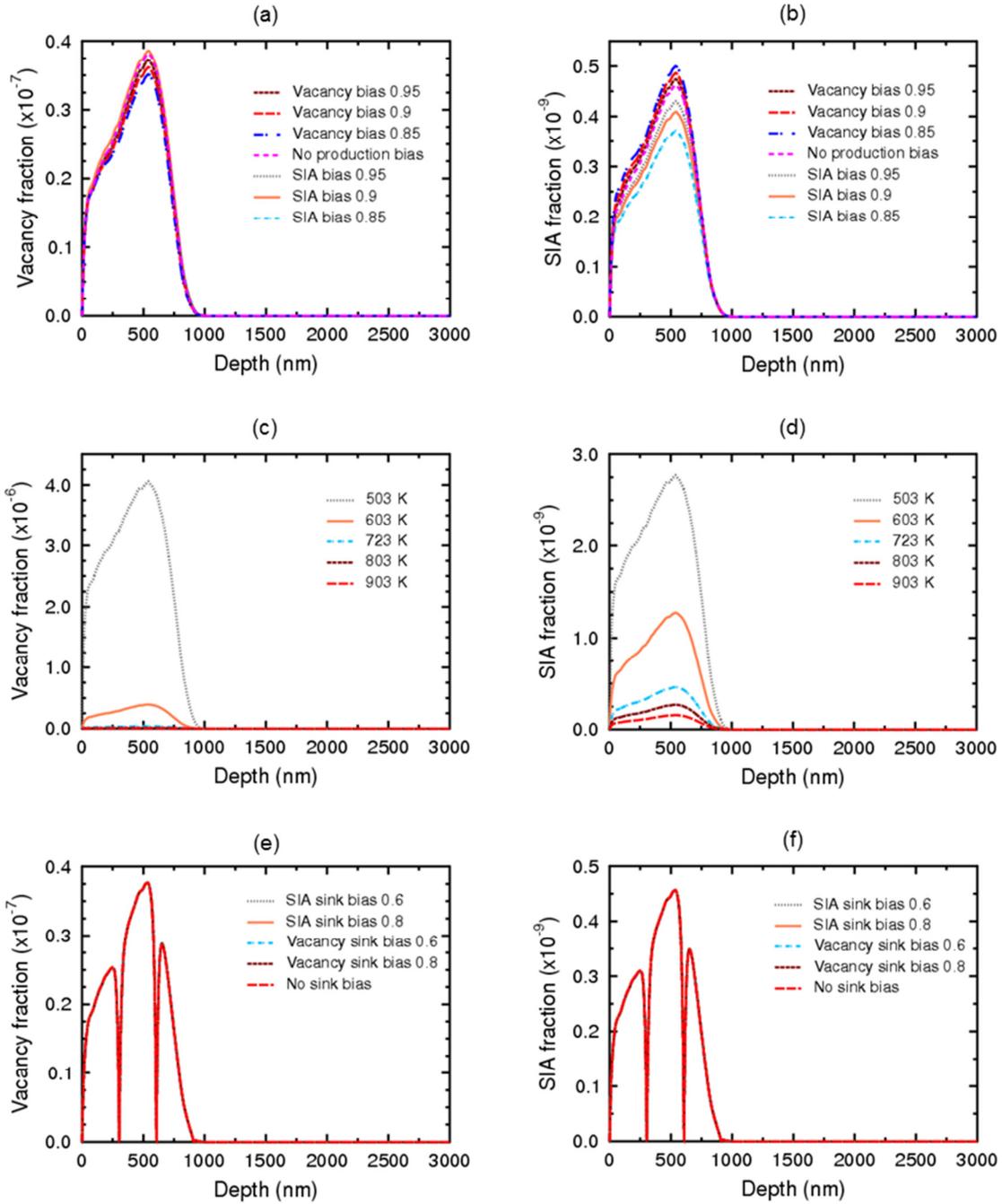

**Figure 13:** Vacancy and self-interstitial atom concentration profiles for pure Fe showing different defect production bias factors (a),(b), temperatures (c),(d), and GB sink bias factors (e),(f). For Fe-14%Cr the vacancy and SIA profiles have smaller concentrations (up to about half as large as for pure Fe), their shapes are similar. Note that all values are given as concentration per lattice site, which we refer to as "fraction" to be consistent with the general notation of the paper.

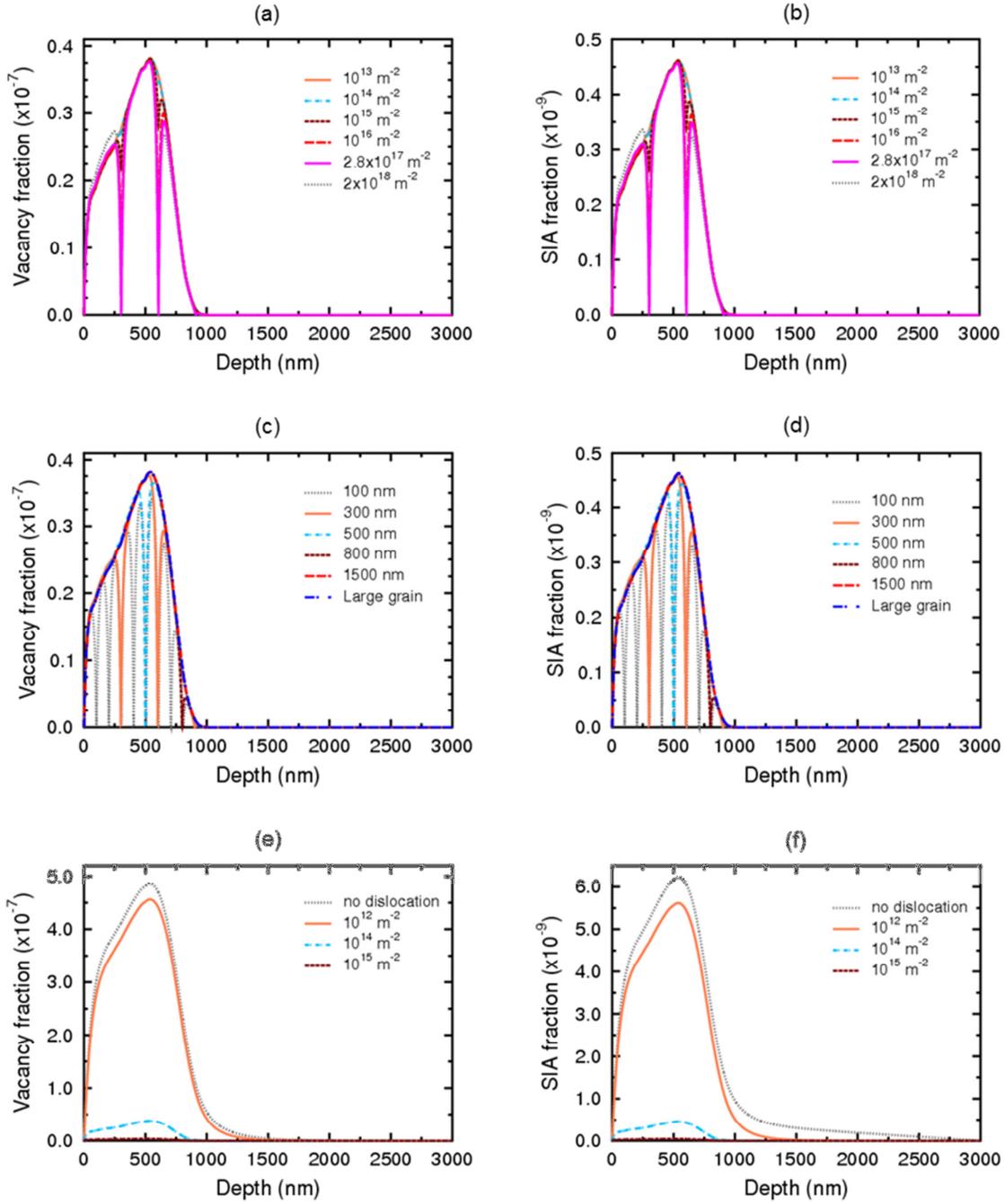

**Figure 14:** Vacancy and self-interstitial atom profiles for pure Fe showing different GB sink strength parameters $\rho_{gb}$ (a), (b), grain sizes (c), (d), and dislocation sink strength parameters $\rho_{DL}$ (e),(f). For Fe-14%Cr the vacancy and SIA profiles have smaller concentrations (up to about half as large as for pure Fe), their shapes are similar. Note that all values are given as concentration per lattice site, which we refer to as "fraction" to be consistent with the general notation of the paper.

## 5. Conclusions

Our rate theory model for defect evolution, which includes the damage and ion implantation profiles, can be used to calculate alloy composition profiles that are useful for understanding and designing heavy ion irradiation experiments.

The SRIM damage and Cr implantation profiles for 1.8 MeV $Cr^+$ irradiating Fe and Fe-14%Cr are quite different from our predicted profiles for the system when the proper kinetics driving diffusion is taken into account. This difference can be seen during the 14 h irradiation with a peak dose of 900 dpa where a significant amount of the implanted Cr migrates from the initial sample and segregates via defect diffusion to the surface and GBs. For instance, SRIM predicts zero Cr concentration up to depth of 100 nm, but our model predicts that Cr concentration will not be lower than 5% due to surface effects. SRIM also predicts 20 at.% Cr in total is deposited locally in the specimen, but our model shows that the concurrent deposition and diffusion limit the local increase to about 6-8 at.% Cr depending on grain size, grain boundary sink strength, and amount of dislocations. These composition changes are important to consider in understanding the implications of ion irradiation for microstructural changes.

The model shows grain boundaries alter the final shape of the Cr profiles and the amount of Cr RIS. Grain size does have an effect on Cr segregation in the Fe-Cr system under ion irradiation, depending on the type of the GB. The Cr stays "locally" near SRIM Cr implantation peak in alloys with small grains because the most of implanted Cr migrates only to the nearest GB sink. The defect absorption at the GB sinks thereby reduces long-range Cr diffusion. Similar effects are seen for increased background dislocation density and increased grain boundary sink strength. Overall, we find a consistent trend that any source which increases average sink strength in the region near the Cr implantation will help maintain the Cr locally near its original implantation location. These results suggest that the potential for significant local Cr enrichment in alloys with average high sink density should be kept in mind when estimating neutron damage using high dose Cr irradiation experiments.

The model also shows that defect production bias, GB sink bias, and temperature do not have significant effects on Cr concentration profiles. Ballistic mixing is also unable to significantly alter the implanted ion profile, which is dominated by radiation enhanced diffusion.

We have shown that the grain boundary type, background dislocation strength, and grain boundary size can influence the shape and content of the Cr profiles after 1.8MeV $Cr^+$ irradiation in both pure Fe and Fe-14%Cr. All Cr segregation in Fe-14%Cr and pure Fe is seen in the damage region and no Cr segregation survives beyond the damage zone because of defect absorption by the high background dislocation. These composition changes are important to consider in understanding the implications of ion irradiation for microstructural changes.

**Acknowledgements**

Financial support for D. Morgan, I. Szlufarska, K. Vörtler, and M. Mamivand was provided by the National Science Foundation (NSF), Division of Materials Research (DMR), Metals and Metallic Nanostructures (MMN), award No. 1105640. L. Barnard acknowledges support from the Rickover Fellowship Program.